# Less is More: Wiring-Economical Modular Networks Support Self-Sustained Firing-Economical Neural Avalanches for Efficient Processing

Junhao Liang[1+], Sheng-Jun Wang[2+], and Changsong Zhou[1,3,4*]

[1]Department of Physics, Centre for Nonlinear Studies and Beijing-Hong Kong-Singapore Joint Centre for Nonlinear and Complex Systems (Hong Kong), Institute of Computational and Theoretical Studies, Hong Kong Baptist University, Kowloon Tong, Hong Kong
[2]Department of Physics, Shaanxi Normal University, Xi'An City, Shaanxi Province, China
[3]Department of Physics, Zhejiang University, 38 Zheda Road, Hangzhou, China
[4]Beijing Computational Science Research Center, Beijing, China
+: equal contribution; *cszhou@hkbu.edu.hk

**Abstract:**

Brain network is remarkably cost-efficient while the fundamental physical and dynamical mechanisms underlying its economical optimization in network structure and activity are not clear. Here we study intricate cost-efficient interplay between structure and dynamics in biologically plausible spatial modular neuronal network models. We find that critical avalanche states from excitation-inhibition balance, under modular network topology with less wiring cost, can also achieve less costs in firing, but with strongly enhanced response sensitivity to stimuli. We derived mean-field equations that govern the macroscopic network dynamics through a novel approximate theory. The mechanism of low firing cost and stronger response in the form of critical avalanche is explained as a proximity to a Hopf bifurcation of the modules when increasing their connection density. Our work reveals the generic mechanism underlying the cost-efficient modular organization and critical dynamics widely observed in neural systems, providing insights to brain-inspired efficient computational designs.

**Key words:** neural network, cost-efficiency, critical avalanche, modular network, mean-field theory

# Introduction

The interplay between structure and dynamics of complex networked systems is always a long-standing topic, covering applications in complex systems from diverse scientific fields. Nowadays, its research and applications in brain and neuroscience are experiencing a rapid growth.

Neurons in human brain form a huge complex dynamical network for efficient functional processing with remarkable cost-efficiency. The principles underlying its efficiency have been actively studied during the past years, either from structure or dynamic aspect.

Brain network is globally very sparse: ~100 billion neurons with ~$10^{14}$ synaptic connections each so that the overall density is ~$10^{-8}$ in human brain [1]. However, the overall low-density connectivity is organized in a hierarchical manner from local circuits, cortical sheets to whole brain connectome [2,3]. Thus, a prominent feature of brain organization is globally sparse with hierarchical relatively dense modular architectures [4–6], which is economical in network wiring since most of the connections are in short-range. There is ample evidence for brain networks to achieve local wiring cost minimization from brain structure [7,8] and a trade-off between global wiring cost and processing efficiency [9,10].

Brain activities consume low energy power of only about 20W, remarkably energy-efficient when compared to digital computers [11]. Dynamically, the irregular and sparse firing of neurons [12] can be collectively organized as oscillations and critical avalanches across different scales [13–16]. Such 'scale-free' dynamic activities are originally explained by critical branching theory [13], where critical avalanches emerge near the transition point between a silent and an overactive phase. Later experimental evidence [14,17] supports that the transition point between an asynchronous and a synchronous phase better explains the observed critical avalanches, especially in terms of the existence of different critical exponents [14,18,19] satisfying scaling relations [20]. Functionally meaningful avalanche dynamics in critical synchronous transition states enable neurons to fire with low rate [21]. Since cortical metabolic energy usage is dominated by action potentials and synaptic transmission [22–26], avalanche dynamic is also energy economical to maintain the sustained spontaneous (resting) state which consumes the majority of brain metabolic cost [27]. Finally, critical states are functionally beneficial to provide a broad dynamic range to stimulations [28,29] and thus a sensitive standing-by state to respond to constantly changing environments for the brain [30].

Though it is recognized that metabolic cost is a unifying principle governing neuronal biophysics [31], the fundamental mechanism underlying the economical interaction between structures and dynamic modes at the neural circuit level is not well understood. Specifically, how the modular network structure and critical dynamics jointly achieve structural and dynamical optimization for energy-efficient processing? Deciphering these mechanisms are also important for developing brain-inspired efficient computing. Here we address these questions with biologically realistic neural dynamic model of excitation-inhibition (E-I) balanced [32,33] spiking neuronal networks, clustered on two-dimensional (2D) space to represent the resting-state dynamics on a cortical sheet composed of micro-columns. Interestingly, when rewiring the initial globally sparse random network into locally dense modular networks, the firing rates decreases, and the self-sustained dynamics changes from asynchronous state to critical avalanche state and the response sensitivity to weak transient external stimuli is greatly enhanced. Theoretically, we reveal the enhanced response of neurons by clustered firing and elucidate the dynamic transition via a Hopf bifurcation induced by denser connections within modules during rewiring, through a novel mean-field theory. Overall, our integrative study of cost-structure-dynamics-function relationships in neural networks elucidates that locally dense connectivity under E-I balanced dynamics appears to be the key "less-is-more" solution to achieve cost-efficient organization.

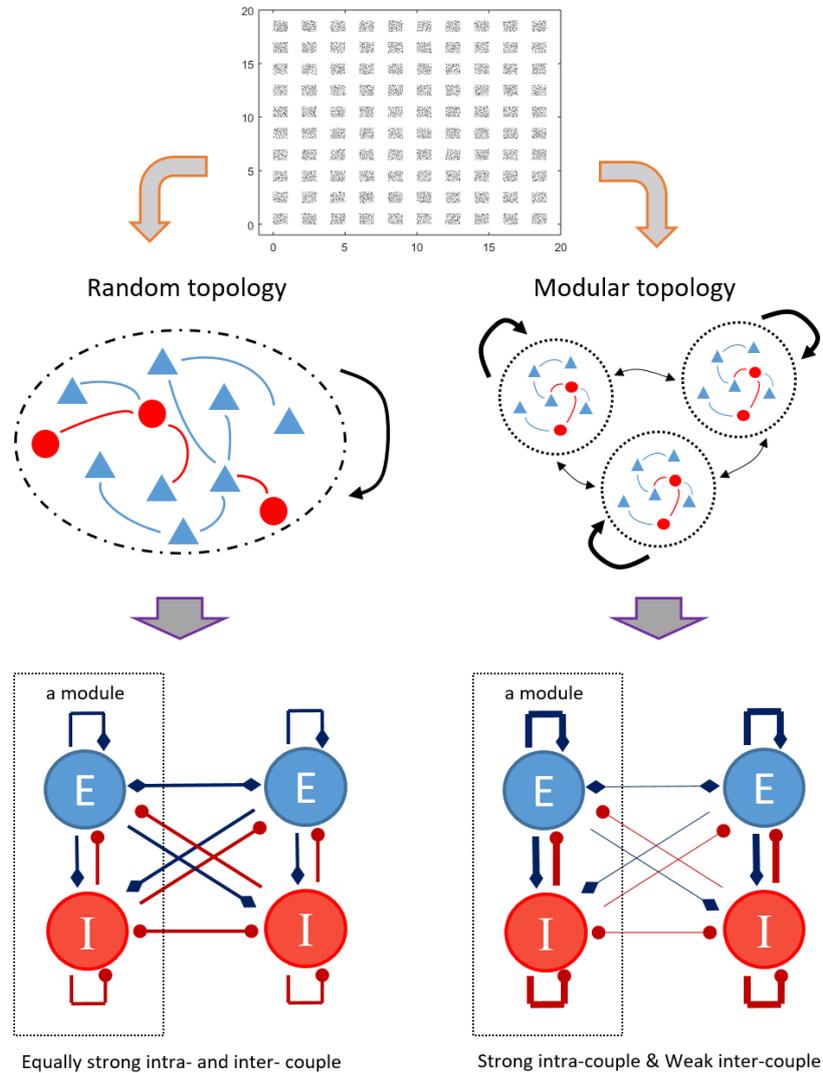

FIG.1. Diagram of the model neural network architecture. Neurons are placed in separated square modules in 2D space (top), mimicking the cortical sheet. Two network wiring patterns are distinguished: globally random topology (left) with equally dense intra- and inter-module coupling and modular topology (right) with dense intra-module coupling but sparse inter-module coupling. In this illustration of the model, the strength of coupling denotes the number of connections.

Results

**Dynamic transition from random network to modular network**

We study a model of $N = 5 \times 10^4$ neurons spread on a 2D plane. Considering that other tissue like vessels may separate micro-columns of neurons, neurons are randomly placed on $N_m = 100$ square regions (modules). Modules are separated by blank space (Fig. 1 top) and each of the modules contains 500 neurons (80% excitatory and 20% inhibitory). Initially, we construct a random network (RN) by randomly connecting each neuron pair with a probability $P_c$ ($P_c = 0.0017$ in the main text). To build a modular network (MN), inter-modular links are rewired, with a probability $P_r$, into the same module to become intra-modular links. The rewiring method is equivalent to construct small-world networks, an essential feature of brain networks [34]. Here the essential structural property captured in our model is that the network consists of coupled modules embedded in space [3,5,9]. The voltage (membrane potential) $V$ of a neuron in the E-I network is governed by conductance-based (COB) leaky integrate-and-fire (IF) dynamic [35],

$$\tau \frac{dV}{dt} = (V_{rest} - V) + g_{ex}(V_E^{rev} - V) + g_{inh}(V_I^{rev} - V). \tag{1}$$

with $\tau, V_{rest}, V_E^{rev}, V_I^{rev}$ being the membrane time constant, resting (leaky) potential, excitatory and inhibitory reversal potential, respectively. When a neuron receives a spike from an E, I neuron, its E, I conductance $g_{ex}, g_{inh}$ is changed as $g_{ex} \to g_{ex} + \Delta g_e$, $g_{inh} \to g_{inh} + \Delta g_i$, respectively, followed by exponentially decay, $\tau_d^E \frac{dg_{ex}}{dt} = -g_{ex}$ and $\tau_d^I \frac{dg_{inh}}{dt} = -g_{inh}$. The network does not receive other external inputs. To launch the network activity, a Gaussian white noise (GWN) is added to Eq. (1) in the initial 200 ms and then removed. Then we study the properties of its self-sustained dynamics without any external inputs. Details of model parameters and simulation methods are provided in Supplementary Notes II. Neural Dynamics.

As the initial RN is rewired into MN, it is interesting to find that the spike-generating and spike-transmission costs of the network are significantly decreased by orders of magnitude (Fig. 2(a)). Here, the spike-generating cost is defined as the average firing rate, and the spike-transmission cost of a neuron is defined as the product of its firing rate and the total length of its outgoing synapses (the average spike-transmission cost shown in Fig. 2(a) is normalized by the value of RN). These two measures for running costs mimic the energy cost for generating spikes and transmitting spikes [36,37]. Besides, since many links become local short-ranged connections, the normalized wiring cost (defined as the total Euclidian length of all links, normalized by the value of RN) is decreased by orders of magnitude and the connection density within modules increases, approaching to a value $p_c \approx 0.17$ ($N_m$ times of the whole network density $P_c$, refer to Eq. (4) below). Such smaller wiring length is desirable as the reduced membrane areas of the fibers can reduces metabolic cost and can reduce the transmission delay (although synaptic delay is not considered in our model for simplicity). The wiring cost reduction is more pronounced for larger networks with more modules (see Fig. S1 and detailed analysis in Supplementary Notes I. Network Setting). Thus, modular network structures reduce both the wiring and running cost.

The initial large and sparse RN can self-sustain asynchronous activity without external input, giving a sustain probability $P_{sus} = 1.0$, which maintains as the network is rewired into MN (Fig. 1(a)).

This self-sustained activity [38] resembles the resting states of brain and thus may play functional role. The results are similar when the overall connection density $P_c$ changes (Fig. S4). However, a denser MN (larger $P_c$) with too weak inter-modular connections ($P_r \to 1$) may not maintain self-sustained activities (Fig. S4). This breakdown of the self-sustainability can be understood later by dynamic analysis of a separate module.

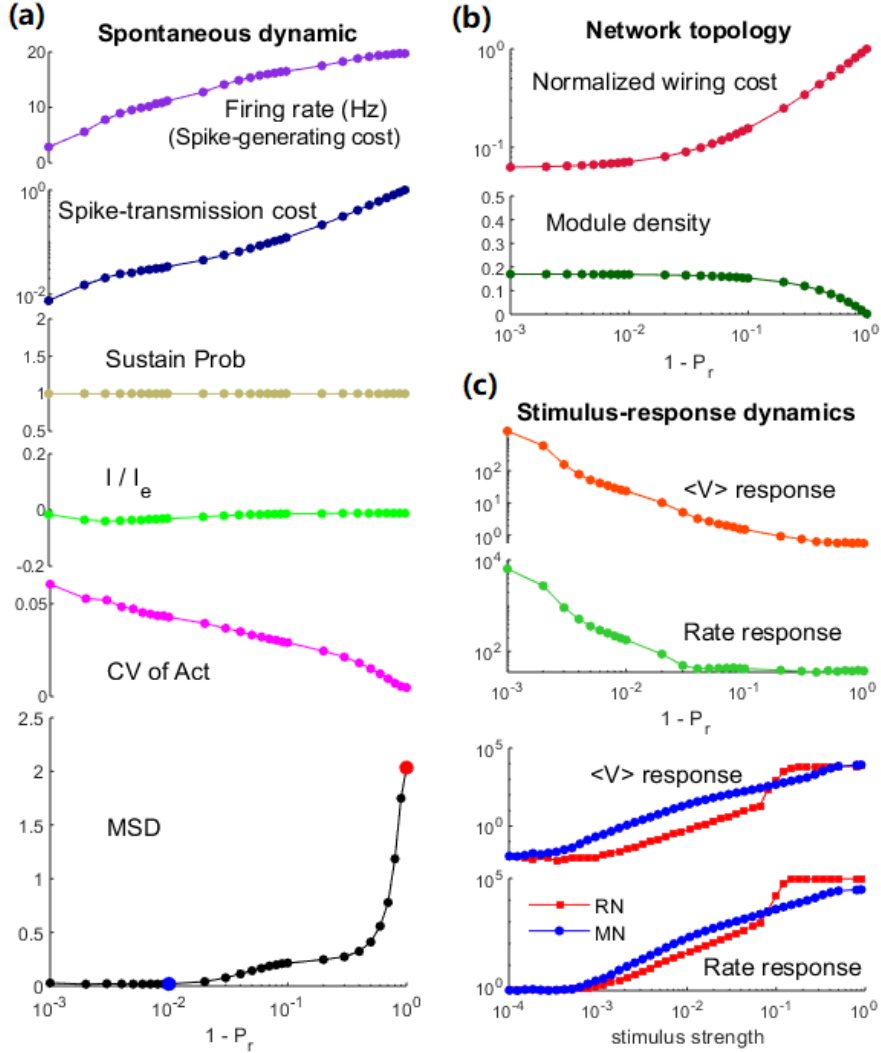

FIG.2. Wiring-economical modular networks support firing-economical avalanches and greatly enhance response sensitivity. (a) Spontaneous dynamic properties of the network during rewiring. From top to bottom: the average firing rate representing spike-generating cost; spike-transmission cost; self-sustained probability $P_{sus}$; current ratio $I/I_e$; CV of activity; and the MSD of avalanche distribution from power-law function. (b) Topological properties of the network during rewiring: normalized wiring cost of the whole network and connection density within a module. (c) Stimulus-response properties. The upper two panels show response size of membrane potential and firing rate during rewiring, where the stimulus strength is 1%. The lower two panels compare the response of RN ($P_r = 0$) and MN ($P_r = 0.99$) under different stimulus strengths.

Very interestingly, the dynamics modes of networks also co-vary during rewiring. RNs exhibit classical E-I balanced asynchronous state with Poisson-like neuronal spiking [32]. We measure the balance by the net synaptic input current rescaled by the excitatory synaptic current ($I/I_E$) averaged over time and neurons. It maintains around 0 when RN is changed into MN (Fig. 2(a)), suggesting the maintenance of overall balance. The asynchronous state in RN has a noisy fluctuation of the mean voltages around an equilibrium value (Fig. 3(a), upper panel). Spikes in MN are clustered yet preserving irregular feature and interrupted by temporally silent periods (refer to Fig. S3(d) for raster plots of the spiking time in a module in MN), exhibiting temporal dynamic variability in mean voltages (Fig. 3(a), lower panel) and it can be measured from CV (coefficient of variability, defined as standard deviation over absolute value of the mean) of the mean voltages of modules in each millisecond (Fig. 2(a)).

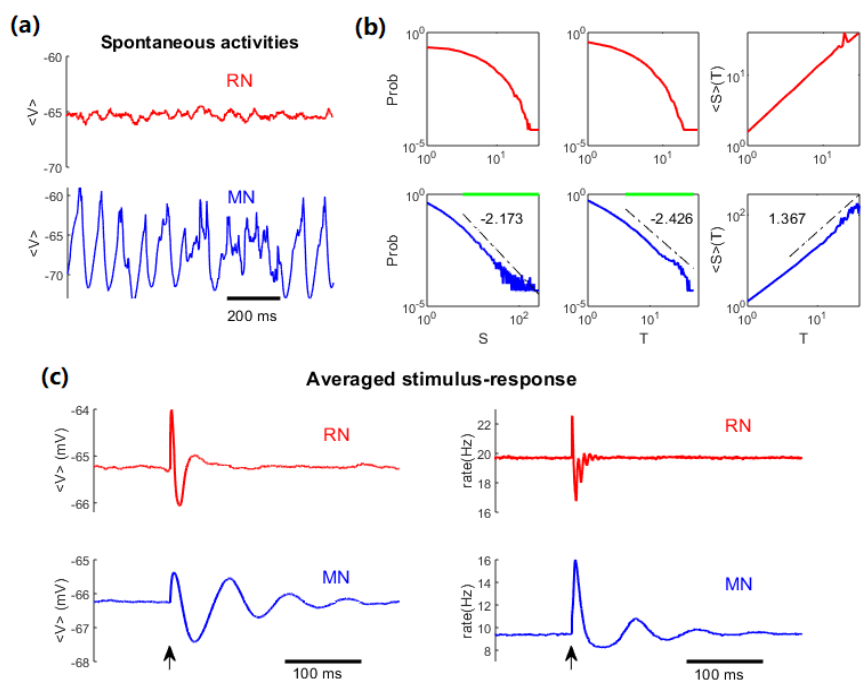

FIG.3. Dynamical comparison between RN ($p_r = 0$) and MN ($p_r = 0.995$). (a) Spontaneous activity (the average membrane potential) in a module. (b) The distributions of avalanche size $S$, avalanche duration $T$, and the average size $\langle S \rangle$ given the duration $T$. The upper/lower rows are results for RN and MN respectively. The green lines on top of the avalanche size and duration distribution under critical dynamics indicate the ranges of estimated power-law distributions. (c) Trial-averaged mean membrane potential (left) and mean firing rate (right) of a module, with transient stimuli (strength 1%) applied at the time marked by the arrows.

Importantly, MN supports critical neuronal avalanches in modules. Here, the time bin for measuring avalanches is the average inter-spike interval (ISI) of the merged spiking train [13] in a module. When rewiring RN into strongly MN (e.g., $P_r$=0.995), the avalanche size and duration distribution of a module changes from exponential-decay to power-law (Fig. 3(b)). Statistical test and estimation of critical exponents are made by accustomed truncation algorithm [39]. Power-law avalanche size and duration distributions: $P(S) \sim S^{-\tau}$, $P(T) \sim T^{-\alpha}$ and $\langle S \rangle(T) \sim T^{1/\sigma vz}$ ($\tau = 2.173$, $\alpha = 2.426$, $\frac{1}{\sigma vz} = 1.367$ with $p$ value $> 0.2$) are found in the truncated ranges, where scaling relation $\frac{\alpha-1}{\tau-1} = \frac{1}{\sigma vz}$ [20] approximately holds (error ~ 0.15). The size distribution is fitted into a power-law function [39] and its mean square deviation (MSD) from the fitted curve in Fig. 2(a) bottom shows that modules in MN with $P_r \geq 0.99$ has avalanches with power law distributions, exhibiting features of criticality. Other measurements of avalanche in MN associated with the threshold of the average membrane potential are presented in Fig. S5, which also exhibit power-law distribution. This transition from asynchronous spiking to critical avalanches dynamics is the approach to a continuous synchronous transition point, as seen from the increase of CV of activity (Fig. 2(a)). The self-sustained activity of coupled modules in the critical states provides the ideal scheme that networks can work with low firing rate. The reduced firing rate at criticality is the feature of the critical synchronous transition model [21], whereas traditional branching process model does not exhibit this property — the firing rate of branching processes at critical states should be larger than that at subcritical states.

Critical states also induce greatly enhanced response sensitivity to transient stimuli. Here, the stimulus is modeled by raising the voltage $V$ to $-40mV$ of a proportion $x$ of the neurons in all modules. We call $x$ the stimulus strength. These neurons driven above the firing threshold emit spikes immediately, similar to optogenetic stimulation in experiments. The response sensitivity of a system can be reflected by the returning process of a signal to its baseline value after a transient perturbation. We measure the response in membrane potential and firing rate of the network modules. The size of response is defined as $\int_{t_0}^{t_0+T} |f(t) - f_b| dt$, which is the area between the signal $f(t)$ (the trial-average voltage or firing rate of the network) and its resting value $f_b$, within a window of $T = 250ms$, beginning from stimulus onset at $t_0$ (see also details in Supplementary Notes II. Neural Dynamics). As shown by the stimulus-induced trial-average voltage and firing rate of a module (Fig. 3(c)), the response of MN is much larger and pronounced than that of RN. Interestingly, MN shows a stronger damped oscillation-like response pattern, which is a characteristic of the event-related potential in electroencephalogram signal of brain's response to stimuli [40]. Importantly, both in membrane potential and in firing rate, MN with critical avalanches exhibits response sensitivity much higher than RN with asynchronous spiking activity for small stimulus strength (Fig. 2(c)). Furthermore, we check the dynamical range, defined as $\Delta = \ln(x_{0.9}/x_{0.1})$ where $x_{0.9}, x_{0.1}$ are the stimulus strengths that induce 90% and 10% response between the minimum and maximum values in logarithmic scale, as in [28,29]. The dynamical ranges of MN ($\Delta_V = 6.28$, $\Delta_r = 5.47$) are greater than that of RN ($\Delta_V = 4.25$, $\Delta_r = 4.25$).

The above structure-dynamics relationships are robust with respect to the overall connection density $P_c$ (see Fig. S4) and also hold in an extended modeling procedure where the number of inter-modular links decay with distance (see Fig. S6). In a word, MN can support cost-efficient critical dynamical modes with greatly enhanced response sensitivity to encode variable input strength, whereas globally sparse RN is both costly in architecture and in running, and cannot properly respond to weak input signals.

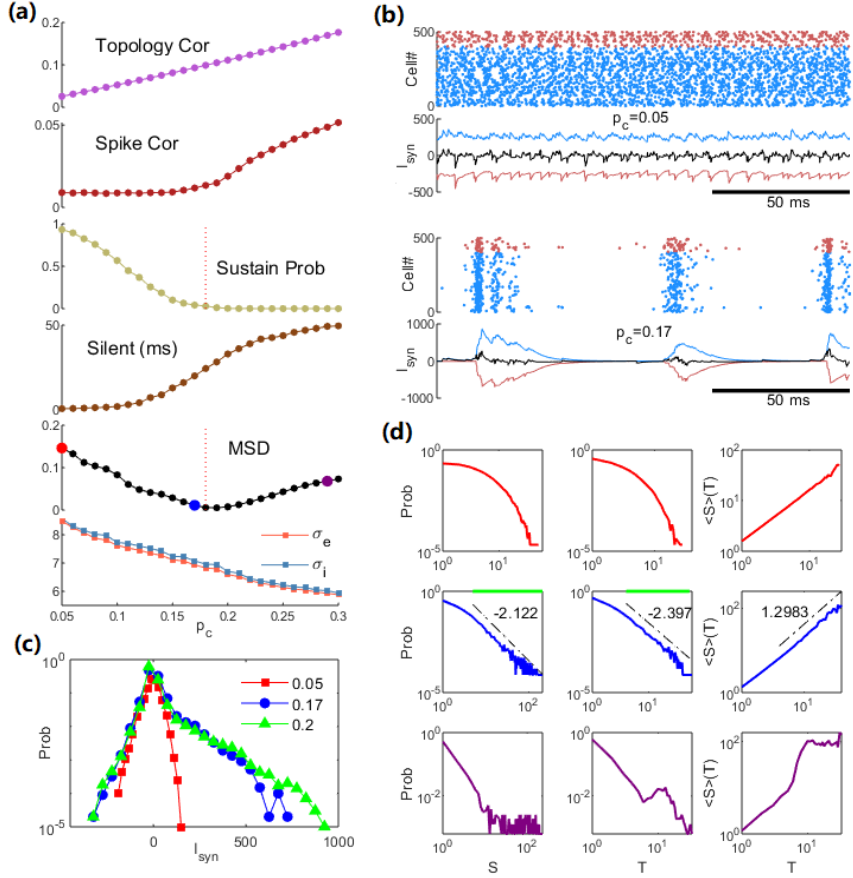

FIG. 4. Spontaneous dynamic of a separate module. (a) The change of properties versus the density $p_c$. From top to bottom: topological correlation; spike correlation; sustained probability; averaged maximum silent period; the MSD of avalanche distribution from power-low fitting function and the numerically estimated effective parameters $\sigma_E$, $\sigma_I$ through Eq. (3). (b) The raster plot of spiking time (blue, red for E, I cells) and the synaptic currents (blue, red, black for E, I, net current) received by a neuron. The upper and lower pannels are for $p_c = 0.05$ and 0.17, respectively. (c) The distribution of net synaptic current received by a neuron for $p_c$=0.05, 0.17, and 0.20. (d) Avalanche distributions (as Fig. 3(b)) for $p_c = 0.05$ (red), 0.17 (blue) and 0.29 (purple).

**Structural correlation and dynamic transition of a single separate module**

The key features in the structure-dynamics relationship can be understood from an isolated module separate from the whole network, but subjected to a background excitatory Poisson input train with rate $r_{in}$. Here, we use $r_{in} = 50\ Hz$, to approximate the weak input received by a neuron from other modules in the highly rewired region in the original network. As the rewiring probability $P_r$ increases, local connection within a module become denser (Fig. 1 and Fig. 2(b)). For a separate module, as its connection density $p_c$ increases, neurons tend to have more common neighbors in the module, the common signal received by a pair of neurons becomes stronger, and their output spikes can be more correlated, as shown in Fig. 4(a).

This correlation in spiking changes the internal interactions in the network. Fig. 4(b) shows the synaptic current of a randomly selected neuron. For low density, the net input current fluctuates slightly around zero due to strong E-I balance (Fig. 4(b), upper panels), and the distribution is close to a normal distribution (Fig. 4(c)). With higher density where spike correlation becomes prominent, correlated excitatory spikes induces quick activation of the network, followed by the activation of inhibitory neurons after an effective delay (due to slower inhibitory synaptic time) and then the activity is depressed. Thus, the net current exhibits oscillations around zero (thus the network maintains E-I balance on average), as shown in Fig. 4(b) lower panels, and its distribution has a large tail at the positive side (Fig. 4(c)). The dynamic pattern is an alternation between synchronized firing and quiescent state with no spikes (Fig. 4(b), lower panels).

Furthermore, as the module becomes denser, the ability of self-sustaining activity of the module decreases (Fig. 4(a)). Here, the self-sustainability is tested by tuning off the external inputs, i.e. letting $r_{in} = 0$ after initial 200ms. The activity almost cannot sustain when $p_c > 0.17$, which is the module density in the original MN when $P_r \to 1$ (Fig. 2(b), also see Eq. (4) below). The weaker sustainable ability of denser network results from clustered firing dynamic mode (Fig. 4(b), lower panel). The silent period during which no neuron fires increases with $p_c$ (Fig. 4(a)). If this period is too long, all recurrent inputs drop out and the network activity dies out since there is no external driving.

Under fixed weak external background inputs $r_{in} = 50$ Hz, as the density $p_c$ increases, the network dynamics undergoes a transition from asynchronous firing pattern (Fig. 4(b) upper panel) to critical avalanches (Fig. 4(b) lower panel) with reduced firing rate (Fig. 6(b)). The MSD of avalanche size distribution from its best-fitted power-law function in Fig. 4(a) shows a minimal around $p_c = 0.18$, close to the transition point of self-sustainability. Typical avalanche distributions for subcritical, critical and supercritical dynamic modes for $p_c = 0.05, 0.17$ and $0.29$ are shown in Fig. 4(d). Power-law avalanche size and duration distributions: $P(S) \sim S^{-\tau}$, $P(T) \sim T^{-\alpha}$ and $\langle S \rangle(T) \sim T^{1/\sigma vz}$ ($\tau = 2.122$, $\alpha = 2.397$, $\frac{1}{\sigma vz} = 1.298$ with $p$ value $> 0.2$) are found for $p_c = 0.17$, where scaling relation $\frac{\alpha-1}{\tau-1} = \frac{1}{\sigma vz}$ [20] approximately holds (error ~ 0.05).

Extended simulation of a separate module (Fig. S7) shows that this dynamical change is independent of input strength $r_{in}$, while the critical density $p_c$ where critical avalanches emerges depends on $r_{in}$. Thus, the emergence of neuronal avalanches can be understood from the large transient fluctuations in the post-synaptic currents induced by correlation, leading to intermittent activities with lower rates.

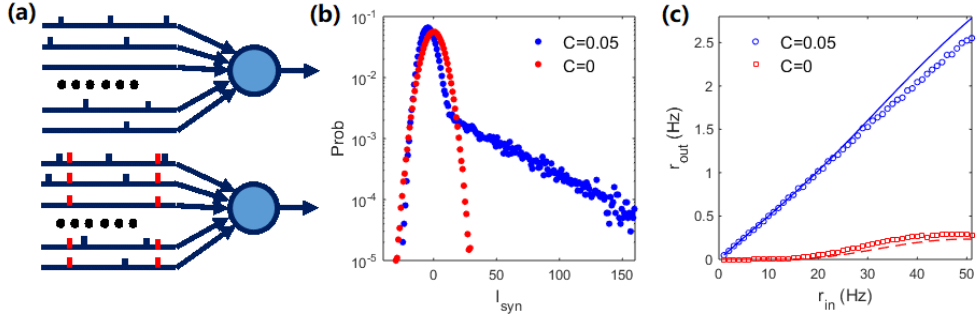

FIG. 5. The simplified illustrative model. (a) The paradigm illustrating a neuron responding to uncorrelated (upper panel) or correlated (lower panel) random spike trains. A Poisson train (labeled as red) is copied into all input neurons to generate correlation among independent spike trains. (b) The distribution of input signal when $r_{in} = 50\ Hz$ when the common spike train is copied into a portion of randomly selected input synapses (see explanation in the text). (c) The response curve between input and output rate. Simulations results (symbols) are compared to theoretical predictions (curves). The red, blue colors are for uncorrelated ($C = 0$) and correlated ($C = 0.05$) input cases.

**Effect of correlation: insight from a simplified model**

To quantitatively illustrate the impact of input correlation on response sensitivity under E-I balanced dynamic, we can consider a simplified model as follows. A single neuron receives spike inputs from other *K=200* Poisson excitatory spike trains. Each of the received spike generates a unit of post-synapse current lasting for $\tau_s = 0.01\ s$. The input signal of the neuron is the summation of these arriving spikes minus a constant equal to the mean current generated by these spikes, in order to mimic the E-I balance. The input correlation is introduced by copying a common Poisson spike train into all input trains [41], see Fig. 5(a) for a paradigm illustration. To construct spike trains with rate $r$ and correlation $C$, the common spike train has a rate $\alpha = Cr$ and independent spike trains have rates $\beta = (1 - C)r$. Assuming a threshold ($\theta = 20$) of the input signal above which the neuron fires a spike, we can numerically obtain the input-output rate response curves (Fig. 5(c)). Compared with independent input trains ($C = 0$), correlation in inputs induces a positive tail in the distribution of input signal (Fig. 5(b)), qualitatively capturing the feature of the IF module (Fig. 4(c)). In the simulation example of Fig. 5(b), the common spike train is copied into a portion of randomly selected input synapses at different time (such that the probability that more synapses receive spikes simultaneously is lower), resulting a decaying positive tail in the distribution of input signal when $C = 0.05$, which quantitatively resembles the observation from the spiking neural network simulations (Fig. 4(c)). We can see that the correlation increases the output rate when the input rate is the same (Fig. 5(c)). Moreover, this simplified model allows an analytic treatment to explain the effect of the correlation on the response rate (see Methods for details). The theoretical results (black dashed line for $C = 0$ and red solid line for $C = 0.05$ in Fig. 5(c)) fit well to the simulation results of this simplified model.

Hence, the correlation in spikes injected from different recurrent synapses improves the responsiveness of neuron. With input correlation and the response sensitivity, each neuron can maintain generating spikes when the overall firing rate is low.

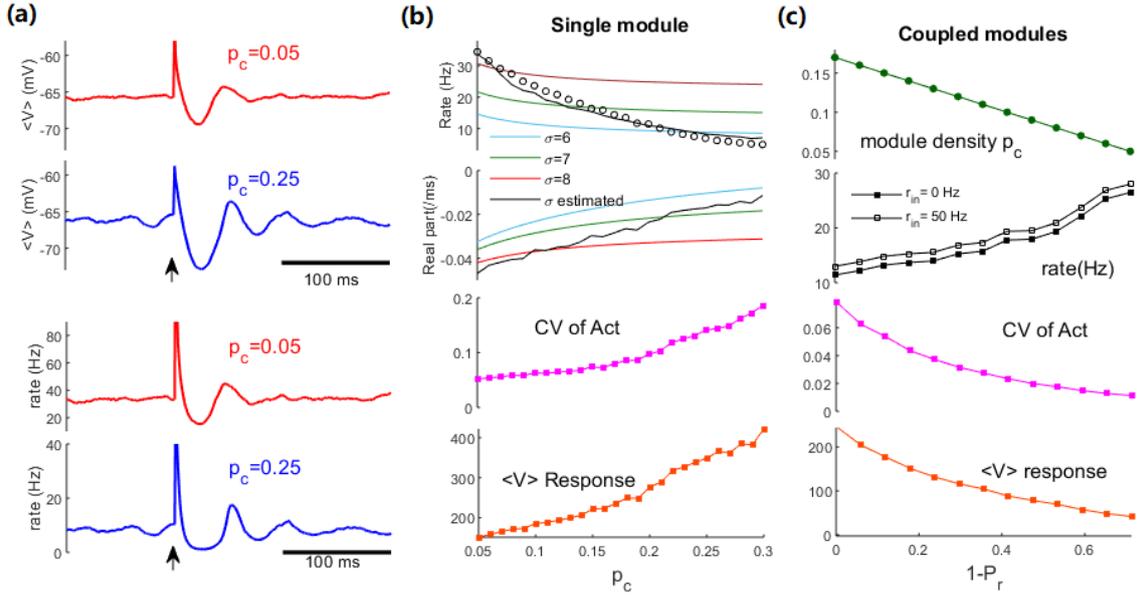

FIG. 6. Mean-field theory to understand the dynamical transition. (a) Examples of the mean membrane potential and mean firing rate evolution of the single-module field equations when $p_c = 0.05$ and $0.25$, with transient stimuli applied at the time marked by the arrows. (b) Properties of the single-module field equations with different density $p_c$. From top to bottom: the firing rate (circles are results from spiking network simulation, curves are results by fixing $\sigma_\alpha = 6,7,8$, and by 'optimal' $\sigma_\alpha$ given in Fig. 4(a)); real part of the eigenvalue of the fixed point; CV of activity; response size of membrane potential. (c) Properties of the coupled field equations with different rewiring probability $P_r$. From top to bottom: the corresponding density within a module; the firing rate; CV of activity; response size of membrane potential.

**Mean-field theory of single module dynamics**

To further understand the dynamical mechanism underlying the transition of dynamics modes together with the reduction of firing rate, we derive the equations of average neural activity in each module and the interaction among the modules by a novel mean-field technique [19] (see Methods and Supplementary Notes II. Neural Dynamics for details). The field equations of a single separate module with connection density $p_c$ receiving excitatory Poisson background input train with rate $r_{in}$ are

$$\begin{cases} \frac{dV_\alpha}{dt} = \frac{V_{rest}-V_\alpha}{\tau} + \left[\tau_d^E g_e\left(r_{in} + \sqrt{\frac{r_{in}}{N_\alpha}}\xi_\alpha(t)\right) + \Phi_E\right](V_E^{rev} - V_\alpha) + \Phi_I(V_I^{rev} - V_\alpha) \\ \frac{d\Phi_\alpha}{dt} = -\frac{\Phi_\alpha}{\tau_d^\alpha} + g_\alpha N_\alpha p_c Q_\alpha, \quad \alpha = E, I \end{cases} \quad (2)$$

where $V_E, V_I$ are the average E, I voltages, $Q_\alpha(t) = 1/[1 + \exp(\frac{V_{th}-V_\alpha}{\sigma_\alpha}\frac{\pi}{\sqrt{3}})]$ are the average firing rate of $\alpha$ neurons, $\Phi_E, \Phi_I$ are the average excitatory, inhibitory synaptic time course received by the neuron and $g_\alpha = \frac{\Delta g_\alpha}{\tau}$. $\xi_\alpha$ are GWN terms. $\sigma_\alpha$ are effective parameters to construct the voltage-dependent mean population firing rate (see Methods for more details). The strong complexity of COB IF dynamics challenges an analytical (self-consistent) estimation of the effective parameters $\sigma_\alpha$ [19]. Taking different fixed $\sigma_\alpha$, the field equations can qualitatively predict the decay of rate with connection density $p_c$ (Fig. 6(b)). To achieve the best prediction, we numerically estimate the effective parameters $\sigma_\alpha$ through the formula

$$\sigma_\alpha = \frac{V_{th} - V_\alpha^{SS}}{\ln((Q_\alpha^{SS})^{-1} - 1)} \frac{\pi}{\sqrt{3}} \qquad (3)$$

from simulations of the single module IF spiking network to numerically obtain the steady-state mean voltage $V_\alpha^{SS}$ and mean firing rate $Q_\alpha^{SS}$ of $\alpha$ neurons. The results of $\sigma_\alpha$ from modules with different density $p_c$ are shown in Fig. 4(a) bottom panel. Under this setting, the field equations can well quantitatively predict the decrease of firing rate as $p_c$ increases (Fig. 6(b)). Note that qualitative prediction can already be achieved by fixing the effective parameters $\sigma_\alpha$ value in Eq. (2) (Fig. 6(b)). Importantly, the field equations reveal that the change of dynamics is associated to a (supercritical) Hope bifurcation. The dominant eigenvalue of the equilibrium in Eq. (2) is complex and its real part approaches zero as $p_c$ increases (Fig. 6(b)). Thus, the firing rate oscillation emerges through approaching to the Hopf bifurcation under noise-perturbation, which induces the critical avalanches [19]. However, the finite-size effect in a small module (500 neurons) hinders the precision of a mean-field theory. Thus, in the spiking IF model the MSD achieve a minimal at around $p_c = 0.17$ (Fig. 4(a)), whereas the field equations do not reach the Hopf bifurcation point and the dynamic is perturbed to bifurcation by noise. Note that a Hopf bifurcation indicates a periodic motion emerges from zero amplitude, corresponding to continuously increasing synchrony as the spiking network (Fig. 2(a)). The CV of activity, measured by the firing rate series of field equations, grows as the increase of $p_c$ (Fig. 6(b)). Finally, response size of voltage computed from the field equations (Fig. 6(b)) also qualitatively predicts the increase of response sensitivity for denser module (examples of $p_c = 0.05$ and $p_c = 0.25$ are shown in Fig. 6(a), compared to Fig. 3(c)). This is because when approaching a bifurcation point, the system will respond more sensitively and will take longer time to damp back to the fixed point after perturbation, a phenomenon called critical slowing down [42].

**Mean-field theory of the modular network**

The above investigation of separated modules with various connection densities under weak external background driving provides understanding of the change of dynamics modes and firing rates with respect to the rewiring probability $P_r$ in the original MN (Fig. 1). First, there is a correspondence between the density in a module $p_c$, and the rewiring probability $P_r$, that

$$p_c = P_c(1 + (N_m - 1)P_r) \qquad (4)$$

(refer to Eq. (S1.5)), as shown in Fig. 6(c) and 2(b). Furthermore, the field equations of the whole MN can be written as (see Methods and Supplementary Notes II. Neural Dynamics for details)

$$\begin{cases} \frac{dV_\alpha^k}{dt} = \frac{V_{rest}-V_\alpha^k}{\tau} + \left[\tau_d^E g_e\left(r_{in} + \sqrt{\frac{r_{in}}{N_\alpha}}\xi_\alpha^k(t)\right) + \Phi_E^k\right](V_E^{rev}-V_\alpha^k) + \Phi_I^k(V_I^{rev}-V_\alpha^k) \\ \frac{d\Phi_\alpha^k}{dt} = -\frac{\Phi_\alpha^k}{\tau_d^\alpha} + g_\alpha N_\alpha P_c[(1+(N_m-1)P_r)Q_\alpha^k + \sum_{l\neq k}(1-P_r)Q_\alpha^l], \alpha = E,I, k=1,\dots,N_m \end{cases}, \quad (5)$$

with $V_\alpha^k, \Phi_\alpha^k, Q_\alpha^k, \xi_\alpha^k$ the corresponding quantities of $\alpha$ neurons in the $k$-th module (see Methods for more details). Thus, the whole MN can be considered as $N_m$ coupled identical neural oscillators. During the rewiring process, the coupling strength between different modules ($\sim 1 - P_r$) reduces whereas self-coupling strength ($\sim 1 + (N_m - 1)P_r$) increases. In this process, although different modules become less affecting each other, the increase of their internal density significantly shapes the dynamic properties of each module as revealed in separated modules (Fig. 4). Here, the effective parameters $\sigma_E, \sigma_I$ to construct $Q_\alpha^k$ in Eq. (5) depend on the rewiring probability $P_r$, through their optimal dependence on $p_c$ in separated modules shown in Fig. 4(a) bottom panel and the relationship between $P_r$ and $p_c$ (Eq. (4)). Numerical results in Fig. 6(c) show that the coupled field equations qualitatively predict the decrease of firing rate, increase of CV of activity and response sensitivity to transient stimuli for increasing rewiring probability $P_r$ as observed in the spiking neural model in Fig. 2. Furthermore, Eq. (5) with $r_{in} = 0$ also predicts the decrease of nonzero firing rate (Fig. 6(c)), which is a qualitative prediction of the self-sustainability during the rewiring (Fig. 2(a)). Note, however, that the mean-field analysis here does not capture the effect of changes of input patterns (e.g. increased input correlation) of a module during the rewiring process of the MN. This is a source of prediction errors that lead to the difference between single-module field equations and a module in the coupled-modules field equations when the latter is constructed by the $\sigma_\alpha$ parameters of the former (refer to Fig. S8). An improvement in the future may be made by assuming oscillatory input in the coupled-modules field model where the oscillatory amplitude increases with the rewiring. To conclude, the mean-field theory predicts the dynamical transition (approaching to a Hopf bifurcation) of a module with increasing internal density and this emergent behavior is maintained for the whole MN with mutually coupled modules when rewiring the inter-modular links to intra-modular links.

## Conclusion and Discussion

In summary, we have unveiled the principle of neural networks allowing cost-efficient optimization in both structure and dynamics simultaneously. There were many studies on either of the sides, considering the optimization of brain network structure [7–9], or the energy-efficient neural dynamics [22–26]. For example, the energy efficient cortical action potentials is facilitated by body temperature [25], and cellular ion channel expression are optimized to achieve function while minimizing metabolic cost of action potential [31]. However, most of previous studies considered the efficiency of network structure (wiring-cost) or the efficiency of dynamics (running-cost) separately. Here, considering both structure and dynamics at the circuit level, we show that wiring-economical modular network can support response-sensitive critical dynamics with much lower running cost while maintaining self-sustainability. This is a notable counterintuitive "less is more" result, because we obtained greatly enhanced functional values with the significant decreases of cost rather than a trade-off between them.

In our model, the efficiency of activity is achieved by critical avalanche states. Different from traditional critical branching region, the critical dynamics in the synchronous transition region simultaneously achieves greater response sensitivity and lower firing rate. Previous studies showed that critical avalanches can appear under various network topology, for instance, scale-free networks with small-world features [43]. Here we show that locally dense while globally sparse MN is the efficient organization of the network structure that enables both the low global wiring cost and the response-sensitive critical dynamics with low running cost. It would be interesting to further explore its dynamic advantages on specific cognitive tasks such as working memory recalling and decision making.

The origin and mechanism of functionally sensitive critical dynamics in neural systems [13–16,28–30] is a long-standing, challenging and controversial topic. Considering the physical mechanism that supports such a co-optimization of structure and dynamic, here we reveal that with increasing topological correlation in E-I balanced network, the spike correlation increases, and so does the fluctuation of the inputs received by neurons. In this case, neurons can be activated by lower firing rate, and the network has higher sensitivity. From the perspective of nonlinear dynamics, these features are captured by a novel mean-field analysis, which reduces the whole modular network into coupled oscillators describing the macroscopic dynamics of each module. We elucidate the dynamical mechanism for producing avalanche as the proximity to a Hopf bifurcation in the mean-field. Close to the bifurcation point, the resulting synchronized spikes in each module are temporally organized as critical avalanches. This stronger collective firing rate variability allows greater computation and coding power [44]. In the highly (yet not totally) rewired MN, the sparse inter-modular connections can provide weak external input to a module from other modules. Meanwhile, as modules are dense enough to be around the response-sensitive critical dynamic states, these weak inputs are enough to maintain the whole MN in self-sustained states with low rates.

In principle, the analytical theory for treating biological plausible COB IF neuronal network is still an open question [45]. Our approximation semi-analytical mean-field technique serves as an effective theory to study the macroscopic dynamics of such realistic networks. It is important to stress that our work put several important features of neural systems into an integrated framework. Spatial embedding of neural circuits under the wring cost constraint gives rise to local dense connections and modular organization [7–9]. The E-I balance is a fundamental property of the neural circuits [32,33]. Collective activities such as critical avalanches and oscillations are pronounced dynamical features of neural networks [13–16,28–30]. Our modeling and theoretical analysis framework reveals intricate interactions among wiring and running costs, modular network topology, critical avalanche dynamic modes and sensitive response to weak stimulations. Thus, it provides an integrative principle for structural-dynamic cost-efficient optimization in neural systems. Our integrative studies with generic network manipulation and novel mean-field theory with realistic neural dynamics can be extended to coupled cortical areas to offer understanding of critical dynamics across the whole brain [46,47] based on a hierarchical modular connectome [48]. Furthermore, spatial networks with connections to only the nearest neighbors can exhibit propagating waves with critical dynamic properties [49]. It would be interesting to generalize our model to such nearest-neighbors coupling scenario and explore the effect of extra sparse long-range connection in such models. This type of models may share similar principles revealed in this study, as in our extended model where short-distance links are dominant (see Fig. S6). The physical principles revealed in our work can guide further development of brain-inspired efficient computing [50].

# Method

### Analysis of the simplified model with correlated inputs

We measure the topological correlation in a RN by the ratio between the number of common neighbors and total number of distinct neighbors of a pair of neurons in the network, that is, $C_{topology} = \frac{(N-2)P_c P_c}{2(N-2)P_c - (N-2)P_c P_c} = \frac{P_c}{2-P_c}$. The spiking correlation (Fig. 4(a)) is measured by the average Person correlation of the spike trains (constructed with window 1 ms) of each neuron in the network.

The relationship between input correlation and response firing rate of the simplified model (Fig. 5(c)) can be obtained as follows. First, in the case without correlation ($C = 0$), the synaptic signal for the target neuron is $s^0(t) = \sum_i [s_i(t) - \tau_s r]$, where $s_i(t)$ is the input of the $i$-th synapse at the time step $t$ and $\tau_s r$ is the mean of the signal such that $\langle s^0(t) \rangle = 0$. We assume the input signal $s^0(t)$ is normally distributed (as in Fig. 4(c)) with the mean $E = 0$ and the variance $\sigma^2(s^0(t)) = K\sigma^2(s_i(t))$. For each synapse, as $s_i = 0$ or $1$ randomly, we have $\langle s_i \rangle = \langle s_i^2 \rangle = \tau_s r$ and the variance is $\sigma^2(s_i) = \tau_s r - (\tau_s r)^2$. Thus, $\sigma^2(s^0) = K[\tau_s r - (\tau_s r)^2]$. The output firing rate is determined by the probability that the input signal is above the threshold $\theta$. Using the error function the firing rate [41] is obtained as

$$r_{out} = \frac{1 - \text{erf}\left(\frac{\theta}{\sqrt{2K\tau_s r (1-\tau_s r)}}\right)}{2\tau_s}, \tag{6}$$

as shown by the red dashed line ($C = 0$) in the Fig. 5(c).

As the correlation of spikes is present in the $K$ input synapses ($C > 0$) when a common spike train is added into all spike trains, the correlated spikes at the time $t$ give the input signal $s^c(t) = K + \sum_i [s_i(t) - \tau_s \beta]$ and its mean strength is $K$, which can active the neuron at time $t$ with probability 1 because $K \gg \theta$. As the rate of the correlated spikes is $\alpha = Cr$, the firing rate of the neuron is

$$r_{out}^c \approx r_{out} + Cr, \tag{7}$$

since the correlated spikes will always induce spiking of the neuron in this simplified model. Here in Eq. (7), the $r_{out}$ is given by Eq. (6) with $r$ replaced by $\beta = (1 - C)r$.

**Mean-field theory of IF neural dynamics**

In this section, we present the outline of the mean-field theory for deriving the field equations Eq. (2) and Eq. (5). More details should be referred to Supplementary Notes II. Neural Dynamics. In our model, for the $i$-th neuron in $k$-th module, we denote its spiking train as $\{t_i^k(n), n \geq 1\}$, its $\alpha$ (E or I) neighbors in the $l$-th module as $\partial_{k,i}^{l,\alpha}$, its voltage as $V_i^k(t)$, its input conductance received from recurrent excitatory, recurrent inhibitory neurons as $GE_i^k(t), GI_i^k(t)$, its external input spike trains (with rate $r_{in}$) and input conductance from external as $\{T_i^k(n), n \geq 1\}$ and $GO_i^k(t)$ (if there are external inputs). Then, the network dynamic equation Eq. (1) can be written in the following more specific form:

$$\begin{cases} \frac{dV_i^k}{dt} = \frac{V_{rest} - V_i^k}{\tau} + [GE_i^k(t) + GO_i^k(t)](V_E^{rev} - V_i^k) + GI_i^k(t)(V_I^{rev} - V_i^k) \\ \frac{dGE_i^k(t)}{dt} = -\frac{GE_i^k(t)}{\tau_d^E} + g_e[\sum_{j \in \partial_{k,i}^{k,E}} \sum_n \delta(t - t_i^k(n)) + \sum_{l \neq k} \sum_{j \in \partial_{k,i}^{l,E}} \sum_n \delta(t - t_i^l(n))] \\ \frac{dGI_i^k(t)}{dt} = -\frac{GI_i^k(t)}{\tau_d^I} + g_i[\sum_{j \in \partial_{k,i}^{k,I}} \sum_n \delta(t - t_j^k(n)) + \sum_{l \neq k} \sum_{j \in \partial_{k,i}^{l,I}} \sum_n \delta(t - t_i^l(n))] \\ \frac{dGO_i^k(t)}{dt} = -\frac{GO_i^k(t)}{\tau_d^E} + g_e \sum_n \delta(t - T_i^k(n)) \end{cases}, \quad (8)$$

where $g_e = \frac{\Delta g_e}{\tau}, g_i = \frac{\Delta g_i}{\tau}$.

Denote $V_E^k = \langle V_i^k \rangle_{i \in k,E}$, $V_I^k = \langle V_i^k \rangle_{i \in k,I}$, $\Phi_E^k = \langle GE_i^k \rangle_{i \in k,E \text{ or } k,I}$ and $\Phi_I^k = \langle GI_i^k \rangle_{i \in k,E \text{ or } k,I}$. We first adopt a diffusion approximation that $GO_i^k(t) \approx \tau_d^E g_e \left( r_{in} + \sqrt{r_{in}} \xi_i^k(t) \right)$, with $\{\xi_i^k(t)\}_{k,j}$ being independent standard GWNs. Thus, $\langle GO_i^k(t) \rangle_{i \in k,\alpha} \approx \tau_d^E g_e \left( r_{in} + \sqrt{\frac{r_{in}}{N_\alpha}} \xi_\alpha^k(t) \right)$, with $\{\xi_\alpha^k(t)\}_{k,\alpha}$ being independent standard GWNs. Then, taking the average $\langle \ \rangle_{i \in k,\alpha}$ to the first equation of Eq. (8), with the decoupling approximation: $\langle [GE_i^k + GI_i^k] V_i^k \rangle_{i \in k,\alpha} \approx \langle GE_i^k + GI_i^k \rangle_{i \in k,\alpha} \langle V_i^k \rangle_{i \in k,\alpha}$ we get the first equation of Eq. (5). Next, the firing rate of the $\alpha$ neurons in the $k$-th module can be approximated as $Q_\alpha^k(t) = \langle \sum_n \delta(t - t_j^k(n)) \rangle_{j \in k,\alpha} = 1/[1 + \exp(\frac{V_{th} - V_\alpha^k}{\sigma_\alpha} \frac{\pi}{\sqrt{3}})]$ [19]. This form essentially captures the sub and supra threshold microscopic dynamics of a spiking network, that is, $Q_\alpha^k(t)$ represents the proportion of $\alpha$ type neurons that spike between $t$ and $t + \Delta t$ ($\Delta t$ is an infinite small quantity) as well as the mean firing rate of $\alpha$ type neurons at time $t$ with unit per ms. Here, $\sigma_\alpha$ are effective parameters to construct the voltage-dependent mean population firing rate. Note that this approximation scheme based only on the first-order statistics neglects several factors that affect the accurate firing rate, including higher order statistics, noise correlation and refractory time. Thus, it does not have an analytical form and $\sigma_\alpha$ should be estimated numerically.

Under mean-field approximation, we have $\langle \sum_{j \in \partial_{k,i}^{l,\alpha}} \sum_n \delta(t - t_j^l(n)) \rangle_{i \in k,E \text{ or } k,I} = n_\alpha^{kl} Q_\alpha^l(t)$, where $n_\alpha^{kl}$ is the average number of $\alpha$ neighbors in the $l$-th module of a neuron in the $k$-th module. Thus, $n_\alpha^{kl} = p^{kl} N_\alpha$, where $p^{kl}$ is the connection probability from module $l$ to module $k$ that

$$p^{kl} = \begin{cases} P_{intra} = P_c(1 + (N_m - 1)P_r), k = l \\ P_{inter} = P_c(1 - P_r), \ k \neq l \end{cases}. \tag{9}$$

Taking $\langle \ \rangle_{i \in k, E}$ or $\langle \ \rangle_{i \in k, I}$ to the second and third equations of Eq. (8), we get the second equation of Eq. (5), which finishes the construction of the coupled field equations.

In the limit of $P_r \to 1$ (all rewired), modules are almost separated. Let $P_r = 1$ in Eq. (5) and we get the field equations corresponding to one separate module with additional external excitatory inputs, i.e. Eq. (2), with $p_c = P_c N_m$ being the connection density of the module.

# Acknowledgement

This work was supported by the Hong Kong Baptist University (HKBU) Strategic Development Fund, the Hong Kong Research Grant Council (GRF12200217), the HKBU Research Committee and Interdisciplinary Research Clusters Matching Scheme 2018/19 (RC-IRCMs/18-19/SCI01) and the National Natural Science Foundation of China (No. 11975194 and 11675096). This research was conducted using the resources of the High-Performance Computing Cluster Centre at HKBU, which receives funding from the Hong Kong Research Grant Council and the HKBU.

Supplementary Materials:

Less is More: Wiring-Economical Modular Networks Support Self-Sustained Firing-Economical Neural Avalanches for Efficient Processing

Junhao Liang, Sheng-Jun Wang, and Changsong Zhou

**This file includes:**

Supplementary Notes I. Network Setting
Supplementary Notes II. Neural Dynamics
Supplementary references
Figures. S1 to S8

**Supplementary Notes I. Network Setting**

The whole system consists of $N$ neurons, 80% are excitatory neurons and 20% are inhibitory neurons. To model a local cortical surface, neurons are placed on square regions which are separated by blank space on a two-dimensional (2D) plane, as illustrated in Fig. 1 top panel in the main text. To measure the distance between different neurons, we first assume the length of each square and the width of the blank space are set as 1 (Fig. 1 top panel). Each neuron $i$, with the coordinate of its position $(x_i, y_i)$, is randomly distributed in each square. In all cases, we assume each module (square) consists of $S_m = 500$ neurons and the ratio between excitatory and inhibitory neurons in each module is kept as 4:1. The example in Fig. 1 top panel contains $N_m = 10 \times 10$ modules and in this case the whole network size is $N = N_m \times S_m = 50,000$. Throughout the work, the default setting is $N_m = 100$, $S_m = 500$ so that $N = 50,000$ and $P_c = 0.0017$ unless extra specifying.

The random network (RN) is built by connecting each possible neuron pair with a probability $P_c$. To build a modular network (MN), links between modules are rewired, with a probability $P_r$, into square to become intra-modular links. For example, for an inter-module link $i \rightarrow j$ whose source neuron $i$ and target neuron $j$ are in different square modules, we replace the target node $j$ with a randomly selected neuron $k$ in the same module of $i$ (initially a link between $i$ and $k$ is absent). This rewiring makes the connection probability in a module higher than that between different modules, forming modular structure in the network while maintaining connection density of the whole network $P_c$ unchanged, that is, the total number of links on average is $M = NP_c(N-1) \approx N^2 P_c$, which is fixed during the rewiring process.

In the initial random network, the number of intra-module links, denoted by $M_{intra}$, is
$$M_{intra} = S_m(S_m - 1)P_c N_m \approx NS_m P_c. \tag{S1.1}$$
The number of inter-modular links in RN, denoted by $M_{inter}$, is
$$M_{inter} = (N_m - 1)NS_m P_c. \tag{S1.2}$$
In this case, both the intra-module and inter-module connection densities are $P_c$, that is,
$$P_{intra} = P_{inter} = P_c. \tag{S1.3}$$
In the MN with the rewiring probability $P_r$, the number of intra-modular links becomes
$$M'_{intra} = M_{intra} + M_{inter}P_r = [1 + (N_m - 1)P_r]NS_m P_c. \tag{S1.4}$$
Thus the intra-module density is
$$P_{intra} = \frac{M'_{intra}}{N(S_m-1)} \approx \frac{M'_{intra}}{NS_m} = P_c(1 + (N_m - 1)P_r). \tag{S1.5}$$
The number of inter-modular links in MN becomes
$$M'_{inter} = M_{inter}(1 - P_r) = (N_m - 1)NS_m P_c(1 - P_r). \tag{S1.6}$$
Thus the inter-module density is
$$P_{inter} = \frac{M'_{inter}}{(N_m-1)NS_m} = P_c(1 - P_r). \tag{S1.7}$$

The length of a link from neuron $i$ to neuron $j$ is the Euclidean distance between the pair of neurons in the 2D plane $l_{ij} = \sqrt{(x_i - x_j)^2 + (y_i - y_j)^2}$. The normalized wiring cost of the whole network is defined as the summation of the length of all links rescaled by that value in initial RN without rewiring. Thus, the normalized wiring cost of initial RN is 1.

Fig. S1 shows the relation between the normalized wiring cost and the rewiring probability $P_r$. As the rewiring probability $P_r$ increases, the wiring cost decreases. When the value of $P_r$ is larger than 0.99, the normalized wiring cost tends to a constant. The results also hold for different module numbers (Fig. S1). This normalized wiring cost in MN is independent of the overall connection density $P_c$, as shown in Fig. S4 bottom panels where the costs with connection density $P_c = 0.001$, $0.0015$, $0.002$ are presented. This property can be understood by an analytic treatment as follows. We assume that the mean length of the intra-modular links is $l$, whereas the mean length of the inter-modular links is $L$. The normalized wiring cost is
$$\text{cost} = \frac{M'_{intra}l + M'_{inter}L}{M_{intra}l + M_{inter}L} = \frac{(S_m-1)l + S_m(N_m-1)P_r l + S_m(N_m-1)(1-P_r)L}{(S_m-1)l + S_m(N_m-1)L}. \tag{S1.8}$$
Thus, it is independent of the connection density $P_c$ of the network, the same as the results shown in Fig. S4 bottom panels. As the rewiring probability $P_r$ tends to 1.0, the normalized wiring cost is
$$\text{cost} = \frac{(S_m-1)l + S_m(N_m-1)l}{(S_m-1)l + S_m(N_m-1)L} \approx \frac{l}{L}. \tag{S1.9}$$
Furthermore, the mean intra-modular link length $l$ does not change with the size of the 2D plane, while the mean inter-modular link length $L$ increases with the size of plane. Therefore, the normalized wiring cost decreases with the size of the system, as shown in Fig. S1.

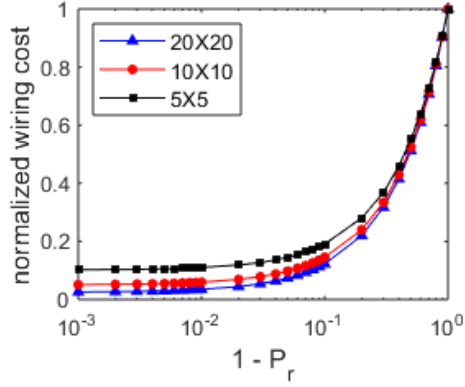

Fig. S1. The normalized wiring cost (rescaled by the initial RN) of MN versus the rewiring probability $P_r$. The number of modules are $N_m = 5 \times 5, 10 \times 10$ and $20 \times 20$ respectively for $P_c = 0.0017$.

## Supplementary Notes II. Neural Dynamics

### 2.1 Spiking dynamics of neurons

In our model, each module has $N_E = 400$ excitatory (E) neurons and $N_I = 100$ inhibitory (I) neurons (totally 500 in a module). For each module, labels 1~400 are E neurons and labels 401~500 are I neurons. For the *i*-th neuron in *k*-th module, we denote its spiking train as $\{t_i^k(n), n \geq 1\}$, its $\alpha$ (E or I) neighbors in the *l*-th module as $\partial_{k,i}^{l,\alpha}$, its voltage as $V_i^k(t)$, its input conductance received from recurrent excitatory, recurrent inhibitory neurons as $GE_i^k(t), GI_i^k(t)$, its external input spike trains (with rate $r_{in}$) and input conductance from external as $\{T_i^k(n), n \geq 1\}$ and $GO_i^k(t)$ (if there are external inputs). Thus, they obey the equations

$$\begin{cases} \frac{dV_i^k}{dt} = \frac{V_{rest} - V_i^k}{\tau} + [GE_i^k(t) + GO_i^k(t)](V_E^{rev} - V_i^k) + GI_i^k(t)(V_I^{rev} - V_i^k) \\ \frac{dGE_i^k(t)}{dt} = -\frac{GE_i^k(t)}{\tau_d^E} + g_e[\sum_{j \in \partial_{k,i}^{k,E}} \sum_n \delta(t - t_j^k(n)) + \sum_{l \neq k} \sum_{j \in \partial_{k,i}^{l,E}} \sum_n \delta(t - t_i^l(n))] \\ \frac{dGI_i^k(t)}{dt} = -\frac{GI_i^k(t)}{\tau_d^I} + g_i[\sum_{j \in \partial_{k,i}^{k,I}} \sum_n \delta(t - t_j^k(n)) + \sum_{l \neq k} \sum_{j \in \partial_{k,i}^{l,I}} \sum_n \delta(t - t_i^l(n))] \\ \frac{dGO_i^k(t)}{dt} = -\frac{GO_i^k(t)}{\tau_d^E} + g_e \sum_n \delta(t - T_i^k(n)) \end{cases},$$

(S2.1)

where $g_e = \frac{\Delta g_e}{\tau}, g_i = \frac{\Delta g_i}{\tau}$. Parameters in simulation are [1]: $\tau = 20ms$, $\tau_d^E = 5ms$, $\tau_d^I = 10ms$, $V_{rest} = -60mV, V_E^{rev} = 0mV$, $V_I^{rev} = -80mV, \Delta g_e = 0.5$, $\Delta g_i = 5$. $\{T_i^n\}$ is the Poisson spiking trains injected to the *i*-th neuron from external whose rate is $r_{in}$ Hz. Spiking reset threshold is $V_{th} = -50mV$. In simulation, we also apply a refractory period 5ms.

In the original networks with 100 modules, there is no external input, i.e. $GO_i^k(t) = 0$. To launch the network activity, a Gaussian white noise (GWN) term, $\xi_i^k(t)$, is added to the first equation of Eq. (S2.1) in the initial 200ms and then removed. It satisfies $\langle \xi_i^k(t) \rangle = 0$ and $\langle \xi_i^k(t)\xi_i^k(s)\rangle = \tau^2 D\delta(t-s)$. $\xi_i^k$ are independent of each other for different $k, i$. Here, we use noise strength $D = 10$. The properties of the self-sustained dynamics is independent of this noise strength.

In the model of a separate module simulation, there is an external input spike train $\{T_i^k(n), n \geq 1\}$ with rate $r_{in} = 50Hz$.

We use 100 realizations of the network in computing the probability of sustained activity $P_{sus}$. If the network still spikes at 1 second after the noise or external input has been removed, it is regarded to exhibit self-sustained activity.

## 2.2 Stimulus-response relation

### 2.2.1 Measure the response sensitivity in spiking networks

In experimental studies on the response of the brain to the cognitive events, it is necessary to average the measured brain activity over trails of experiments. Here, the stimulus method is that $x$ proportion of the randomly selected neurons in all modules are activated by increasing their membrane potential to $-40mV$ (above the threshold) in one simulation step. $x$ is termed the stimulus strength in our study.

We measure the response sensitivity in terms of the membrane potential properties and the spiking properties, using the average membrane potential and firing rate of the network modules. To study the sensitivity, we can consider the returning process of a signal to its baseline value after a transient perturbation.

The collective behavior of a module $k$ responding to additional stimulus can be reflected by its mean membrane potential and its mean firing rate: $\langle V^k(t)\rangle = \frac{1}{s_m}\sum_{i=1}^{s_m} V_i^k(t)$, $\langle Q^k(t)\rangle = \frac{1}{s_m}\sum_{i=1}^{s_m} Q_i^k(t)$ (where $Q_i^k(t)$ is the neuronal firing rate series constructed in 1ms).

Averaged signals $\langle V(t)\rangle = \frac{1}{N_m}\sum_{k=1}^{N_m}\langle V^k(t)\rangle$, $\langle Q(t)\rangle = \frac{1}{N_m}\sum_{k=1}^{N_m}\langle Q^k(t)\rangle$ (averaged over all the modules) are then used as the measured signal $f(t)$. We measure the response size by $\int_{t_0}^{t_0+T}|f(t)-f_b|dt$, the area of the region (colored region in Fig. S2) between the signal curve $f(t)$ and the baseline of the ongoing activity $f_b$. The length for recording the response is $T = 250ms$ after the stimulus onset at $t = t_0$. Results are averaged over 100 realizations of the simulation.

An example of the averaged signal is shown in Fig. S2. In general, the ongoing spontaneous fluctuation is almost eliminated by averaging over realizations, but the response behavior manifests itself in the averaged signal. If the response of the network is weak, the network returns to its baseline ongoing activity quickly, whereas strong response causes a waveform of damped oscillation as illustrated in Fig. S2 (see also Fig. 3(c) in the maintext for the cases of a module in RN and MN).

## 2.2.2 Measure the response sensitivity in the field models

We can apply similar analysis in the field equations simulation. In the single-module field equation Eq. (S2.10), the stimulus is to raise the E, I membrane potential ($V_E$ and $V_I$) above the threshold to $-40mV$. Then, we measure the corresponding area index calculated from the average membrane potential of the module $V = 0.8V_E + 0.2V_I$. In the coupled version of the field equations Eq. (S2.6), we exert the above stimulus in a selected module $k$ and detect the response by the signal of this module as described above.

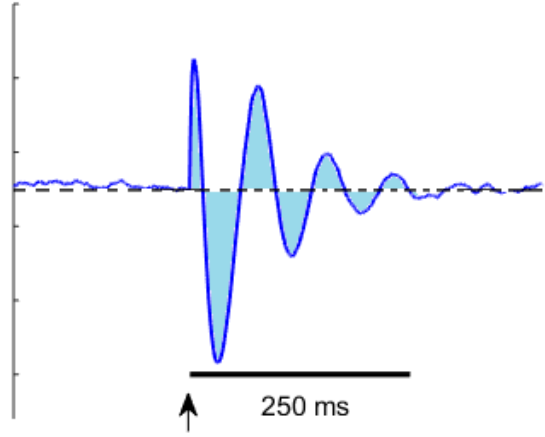

Fig. S2. The illustration of computing the response size. The colored area between the signal curve (the trial-average voltage or firing rate) and its pre-stimulus baseline value (dashed line) in 0~250ms after stimuli is used to measure the response size. Stimuli are applied at the time marked by the arrows. Refer to Fig. 3(c) for the cases of RN and MN in the network model.

## 2.3 Mean-field reduction of neuronal network to coupled neural oscillators

Now we derive the macroscopic field equations corresponding to the spiking network (S2.1). Denote $V_E^k = \langle V_i^k \rangle_{i \in k, E}$, $V_I^k = \langle V_i^k \rangle_{i \in k, I}$, $\Phi_E^k = \langle GE_i^k \rangle_{i \in k, E \text{ or } k, I}$ and $\Phi_I^k = \langle GI_i^k \rangle_{i \in k, E \text{ or } k, I}$ as the average E, I voltage, E, I input conductance of the $k$-module. For external inputs of neuron $i$, we adopt a diffusion approximation that $GO_i^k(t) \approx \tau_d^E g_e \left( r_{in} + \sqrt{r_{in}} \, \xi_i^k(t) \right)$, with $\{\xi_i^k(t)\}_{k,j}$ being independent standard (with zero mean and unit variance) GWNs. Thus, $\langle GO_i^k(t) \rangle_{i \in k, \alpha} \approx \tau_d^E g_e \left( r_{in} + \sqrt{\frac{r_{in}}{N_\alpha}} \xi_\alpha^k(t) \right)$, with $\{\xi_\alpha^k(t)\}_{k,\alpha}$ being independent GWNs. Taking the average $\langle \ \rangle_{i \in k, E}$ and $\langle \ \rangle_{i \in k, I}$ to the first equation of Eq. (S2.1), with the decoupling approximation: $\langle [GE_i^k + GI_i^k] V_i^k \rangle_{i \in k, \alpha} \approx \langle GE_i^k + GI_i^k \rangle_{i \in k, \alpha} \langle V_i^k \rangle_{i \in k, \alpha}$ we get

$$\frac{dV_\alpha^k}{dt} = \frac{V_{rest} - V_\alpha^k}{\tau} + \left[ \tau_d^E g_e \left( r_{in} + \sqrt{\frac{r_{in}}{N_\alpha}} \, \xi_\alpha^k(t) \right) + \Phi_E^k \right] \left( V_E^{rev} - V_\alpha^k \right) + \Phi_I^k \left( V_I^{rev} - V_\alpha^k \right). \quad (S2.2)$$

By the assumption [2], we know that the firing rate of the $\alpha$ neurons in the $k$-th module can be approximated as

$$Q_\alpha^k(t) = \langle \sum_n \delta\left(t - t_j^k(n)\right) \rangle_{j \in k,\alpha} = 1/[1 + \exp(\frac{V_{th}-V_\alpha^k}{\sigma_\alpha^k}\frac{\pi}{\sqrt{3}})] \ . \tag{S2.3}$$

Eq. (S2.3) essentially captures the sub and supra threshold microscopic dynamics of a spiking network, that is, $Q_\alpha^k(t)$ represents the proportion of $\alpha$ type neurons that spike between $t$ and $t + \Delta t$ ($\Delta t$ is an infinite small quantity) as well as the mean firing rate of $\alpha$ type neurons at time $t$ with unit per ms [2]. Here, $\sigma_\alpha^k$ are effective parameters to construct the voltage-dependent mean population firing rate. Note that this approximation scheme based only on the first-order statistics neglects several factors that affect the accurate firing rate, including higher order statistics, noise correlation and refractory time. Thus, $\sigma_\alpha^k$ does not have an analytical form and should be estimated numerically. A complete analytical approach for conductance-based integrate-and-fire neural network is still an open issue [3,4]. Furthermore, the quality of the scheme depends on suitable choices of effective parameters $\sigma_\alpha^k$ (see Eq. (S2.11) below).

Under mean-field approximation, we have $\langle \sum_{j \in \partial_{k,i}^{l,\alpha}} \sum_n \delta\left(t - t_j^l(n)\right) \rangle_{i \in k,E \text{ or } k,I} = n_\alpha^{kl} Q_\alpha^l(t)$, where $n_\alpha^{kl}$ is the average number of $\alpha$ neighbors in the $l$-th module of a neuron in the $k$-th module. Thus, $n_\alpha^{kl} = p^{kl} N_\alpha$, where $p^{kl}$ is the connection probability from module $l$ to module $k$. In our network model, from Eq. (S1.5) and Eq. (S1.7), we have

$$p^{kl} = \begin{cases} P_{intra} = P_c(1 + (N_m - 1)P_r), k = l \\ P_{inter} = P_c(1 - P_r), \ k \neq l \end{cases} . \tag{S2.4}$$

Taking $\langle \ \rangle_{i \in k,E}$ or $\langle \ \rangle_{i \in k,I}$ to the second and third equations of Eq. (S2.1), we have

$$\frac{d\Phi_\alpha^k}{dt} = -\frac{\Phi_\alpha^k}{\tau_d^\alpha} + g_\alpha N_\alpha P_c[(1 + (N_m - 1)P_r)Q_\alpha^k(t) + \sum_{l \neq k}(1 - P_r)Q_\alpha^l(t)] \ . \tag{S2.5}$$

Thus, the field equations of the whole MN are obtained as

$$\begin{cases} \frac{dV_\alpha^k}{dt} = \frac{V_{rest}-V_\alpha^k}{\tau} + \left[\tau_d^E g_e\left(r_{in} + \sqrt{\frac{r_{in}}{N_\alpha}}\ \xi_\alpha^k(t)\right) + \Phi_E^k\right](V_E^{rev} - V_\alpha^k) + \Phi_I^k(V_I^{rev} - V_\alpha^k) \\ \frac{d\Phi_\alpha^k}{dt} = -\frac{\Phi_\alpha^k}{\tau_d^\alpha} + g_\alpha N_\alpha P_c[\frac{(1+(N_m-1)P_r)}{1+\exp\left(\frac{V_{th}-V_\alpha^k}{\sigma_\alpha^k}\frac{\pi}{\sqrt{3}}\right)} + \sum_{l \neq k}\frac{(1-P_r)}{1+\exp\left(\frac{V_{th}-V_\alpha^l}{\sigma_\alpha^l}\frac{\pi}{\sqrt{3}}\right)}], \alpha = E, I, k = 1, \dots, N_m \end{cases} . \tag{S2.6}$$

Since all modules are identical, we have $\sigma_\alpha^k = \sigma_\alpha$ for all $k$. Denote $X^k = \left(V_E^k, V_I^k, \Phi_E^k, \Phi_I^k\right)^T$, we can write the field equations in the vector form as

$$\frac{dX^k}{dt} = F(X^k) + \sum_{j=1}^{N_m} a_{kj} G(X^j) \ , \tag{S2.7}$$

with

$$F(X^k) = \{\frac{V_{rest} - V_E^k}{\tau} + \left[\tau_d^E g_e\left(r_{in} + \sqrt{\frac{r_{in}}{N_E}}\ \xi_E^k(t)\right) + \Phi_E^k\right](V_E^{rev} - V_E^k) + \Phi_I^k(V_I^{rev} - V_E^k),$$

$$\frac{V_{rest}-V_I^k}{\tau} + \left[\tau_d^E g_e\left(r_{in} + \sqrt{\frac{r_{in}}{N_I}}\, \xi_I^k(t)\right) + \Phi_E^k\right](V_E^{rev} - V_I^k) + \Phi_I^k(V_I^{rev} - V_I^k), -\frac{\Phi_E^k}{\tau_d^E}, -\frac{\Phi_I^k}{\tau_d^I}\}^T ,$$

(S2.8a)

$$G(X^k) = \left\{0, 0, \frac{g_E N_E P_c}{1+\exp\left(\frac{V_{th}-V_E^k}{\sigma_E}\frac{\pi}{\sqrt{3}}\right)}, \frac{g_I N_I P_c}{1+\exp\left(\frac{V_{th}-V_I^k}{\sigma_I}\frac{\pi}{\sqrt{3}}\right)}\right\}^T, \quad (S2.8b)$$

$$a_{kj} = \begin{cases} 1 + (N_m - 1)P_r, & k = j \\ 1 - P_r, & k \neq j \end{cases}. \quad (S2.8c)$$

One should notice that the effective parameters $\sigma_E, \sigma_I$ in Eq. (S2.6) should be determined by all other parameters (particular by rewiring probability $P_r$).

For example, if $\sigma_E, \sigma_I$ are independent of $P_r$, then at the deterministic steady-state where $\dot{V}_\alpha^k = \dot{Q}_\alpha^k = \xi_\alpha^k = 0$, one expect that $V_\alpha^k = V_\alpha^l$, $\Phi_\alpha^k = \Phi_\alpha^l$, $Q_\alpha^k = Q_\alpha^l$ for all $k \neq l$ since the identity of different modules. Then, these steady values are solved by the algebraic equations

$$\begin{cases} \frac{V_{rest}-V_\alpha^{ss}}{\tau} + [\tau_d^E g_e r_{in} + \Phi_E^{ss}](V_E^{rev} - V_\alpha^{ss}) + \Phi_I^{ss}(V_I^{rev} - V_\alpha^{ss}) \\ \Phi_\alpha^{ss}(t) = \tau_d^\alpha g_\alpha N_\alpha P_c N_m / [1 + \exp\left(\frac{V_{th}-V_\alpha^{ss}}{\sigma_\alpha}\frac{\pi}{\sqrt{3}}\right)] \end{cases}. \quad (S2.9)$$

Thus, the steady-state firing rate $Q_\alpha^{ss}(t) = 1/[1 + \exp(\frac{V_{th}-V_\alpha^{ss}}{\sigma_\alpha}\frac{\pi}{\sqrt{3}})]$ is independent of $P_r$ if $\sigma_E, \sigma_I$ are independent of $P_r$, which cannot capture the effect of the firing rate reduction during rewiring (Fig. 2 in the main text).

**2.4 Field equations of a single module**

To understand the overall dynamic principle of the modular network, we can focus on analyzing the dynamics of single separate module. In the limit of $P_r \to 1$ (all rewired), modules are almost separated. Let $P_r = 1$ in Eq. (S2.6) and we get the field equations corresponding to one separate module with additional external excitatory inputs:

$$\begin{cases} \frac{dV_\alpha}{dt} = \frac{V_{rest}-V_\alpha}{\tau} + \left[\tau_d^E g_e\left(r_{in} + \sqrt{\frac{r_{in}}{N_\alpha}}\,\xi_\alpha(t)\right) + \Phi_E\right](V_E^{rev} - V_\alpha) + \Phi_I(V_I^{rev} - V_\alpha) \\ \frac{d\Phi_\alpha}{dt} = -\frac{\Phi_\alpha}{\tau_d^\alpha} + g_\alpha N_\alpha p_c / \left[1 + \exp\left(\frac{V_{th}-V_\alpha}{\sigma_\alpha}\frac{\pi}{\sqrt{3}}\right)\right] \end{cases}, \quad \alpha = E, I$$

(S2.10)

where $p_c = P_c N_m$ in the connection density of this module. From this we can know how the dynamic properties of a module depends on its effective connection density $p_c$. In our study here, the effective parameters $\sigma_E, \sigma_I$ are estimated by

$$\sigma_\alpha = \frac{V_{th}-V_\alpha}{\ln((Q_\alpha^{ss})^{-1}-1)}\frac{\pi}{\sqrt{3}}, \alpha = E, I, \quad (S2.11)$$

where $V_\alpha^{ss}$ and $Q_\alpha^{ss}$ are the steady-state average membrane potential and firing rate of E, I neurons in the single separate module and they are estimated through numerical simulation of the module under different connection density $p_c$ (Fig. 4(a) bottom panel). Although qualitative prediction may not depend on the exact value of $\sigma_\alpha$, (see Fig. 6(b)), we adopt Eq. (S2.11) to achieve a better predictive outcome.

## Supplementary references

**Other supplementary figures**

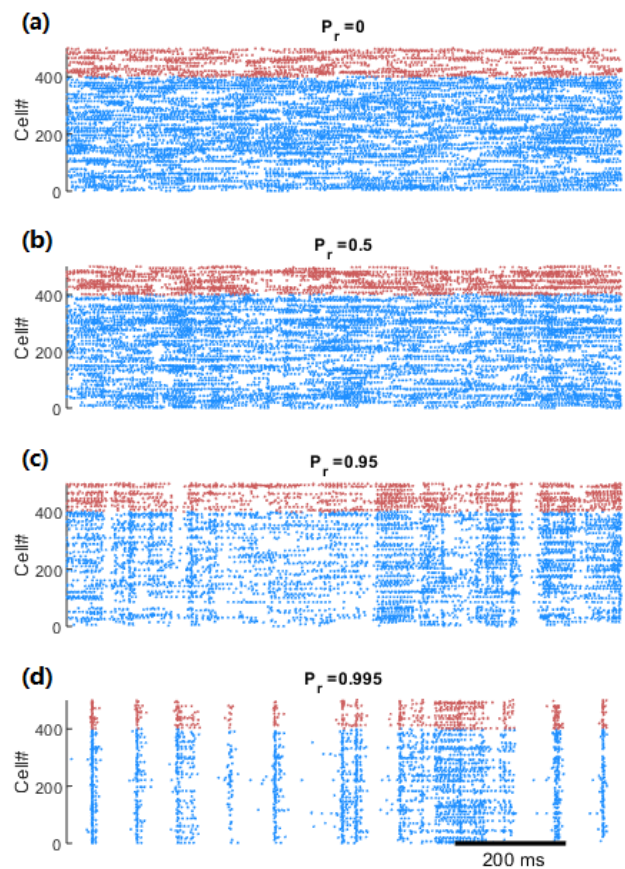

Fig. S3. The raster plot of spike times (blue, red for E, I cells) in a module selected from the whole network. From (a) to (d) the rewiring probability is $P_r = 0, 0.5, 0.95, 0.995$ respectively. As $P_r$ grows, the dynamic mode transitions from asynchronous spiking to intermittent clustered spiking.

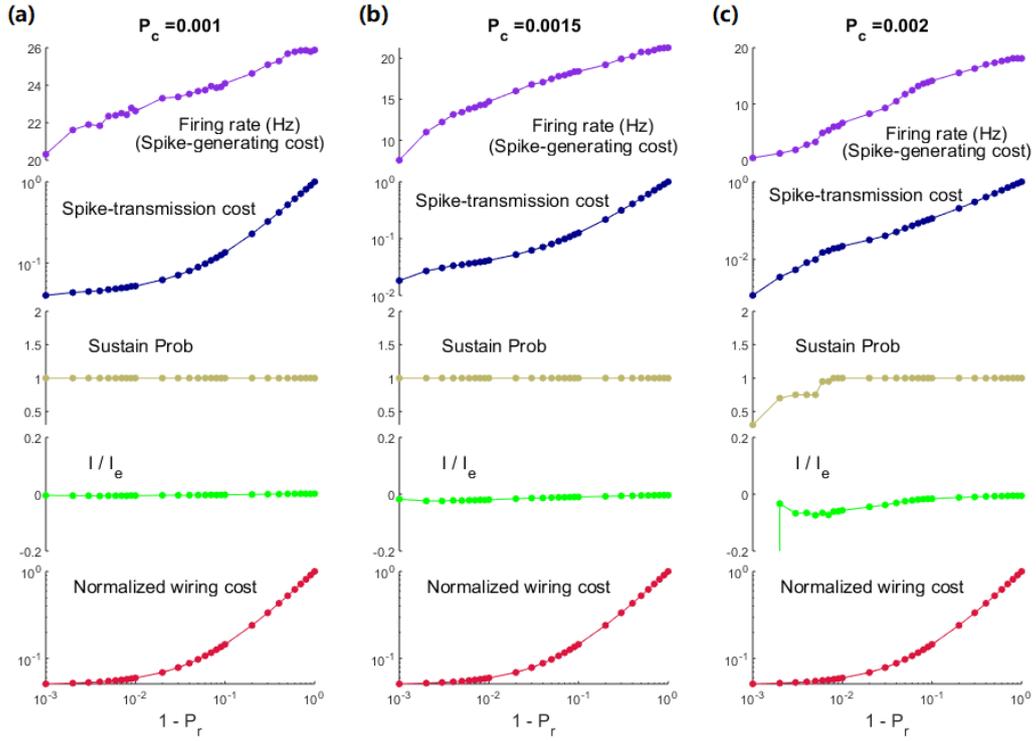

Fig. S4. Comparison of network properties during rewiring under different global density $P_c$. From top to bottom: the average firing rate; spike-transmission cost; sustained probability $P_{sus}$; current ratio $I/I_e$ and normalized wiring cost. (a) $P_c = 0.001$. (b) $P_c = 0.0015$. (c) $P_c = 0.002$. Results in the main text are obtained by $P_c = 0.0017$. Conclusions are the same when $P_c$ is around this value. However, MN (for $P_r \to 1$) may not be able to self-sustain when $P_c$ is larger, as shown in (c). This can be understood from the self-sustain property of a single isolated module (Fig. 4(a)). However, since we do not employ synaptic scaling (i.e. let the synaptic strength decreases with neighbor numbers) in our model, the network cannot maintain E-I balance for too large $P_c$ (data not shown).

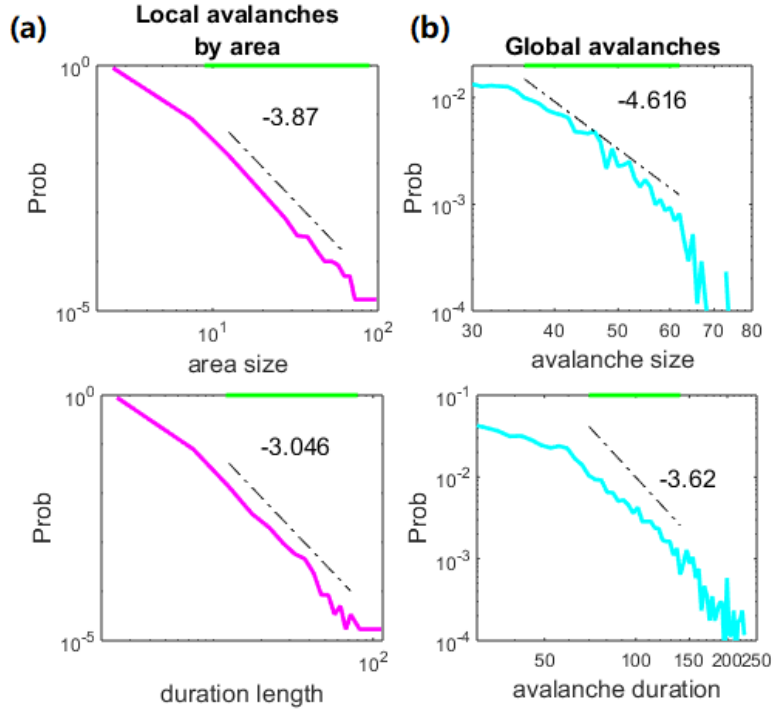

Fig. S5. Additional measurements of avalanches. In the main text, we measure avalanches from neuronal spikes (Fig. 3 and 4). Here, we present other measurements of avalanches using the average membrane potential associated with a threshold $V_0$. (a) Avalanches defined by the above-threshold-event of the average membrane potential $\langle V \rangle$ of a module, as used in [5]. The size of an avalanche is defined as the area below the curve $\langle V \rangle$ and above the threshold in the event. The duration of an avalanche is defined as the time length of the event. The size and duration probability distributions of avalanches defined in this way are shown ($P(S) \sim S^{-\tau}$, $(T) \sim T^{-\alpha}$  $\tau = 3.87$, $\alpha = 3.046$, with $p$ value $> 0.1$). Measures are taken in a module of MN with $P_r = 0.995$, $P_c = 0.0017$. (b) Global avalanche between modules. The threshold events of each module is first defined when the average membrane potential of each module $\langle V^k \rangle$ reaches threshold $V_0$. Then the spread of threshold events in $N_m = 100$ modules is analyzed with method similar to spiking data with a time bin $\Delta t = 3.5 ms$. This defines the mesoscopic avalanches as studied in MEG data [6]. Avalanche size and duration distribution are shown ($P(S) \sim S^{-\tau}$, $(T) \sim T^{-\alpha}$  $\tau = 4.616$, $\alpha = 3.62$, with $p$ value $> 0.1$). Measures are taken in a module of MN with $P_r = 0.99$, $P_c = 0.0017$. In (a, b), a threshold $V_0 = -60 mV$ is used. The threshold is approximately the mean ($-66.7 mV$) plus 2 times of the standard deviation ($2.9 mV$) of $\langle V \rangle$, as in previous studies. The green lines on the top indicate the ranges of estimated power-law distributions. The results suggest that our model can also exhibit features of critical avalanches in mesoscopic scale. Their properties deserve further study, for example, in terms of critical exponents and in association to the underlying spiking avalanches.

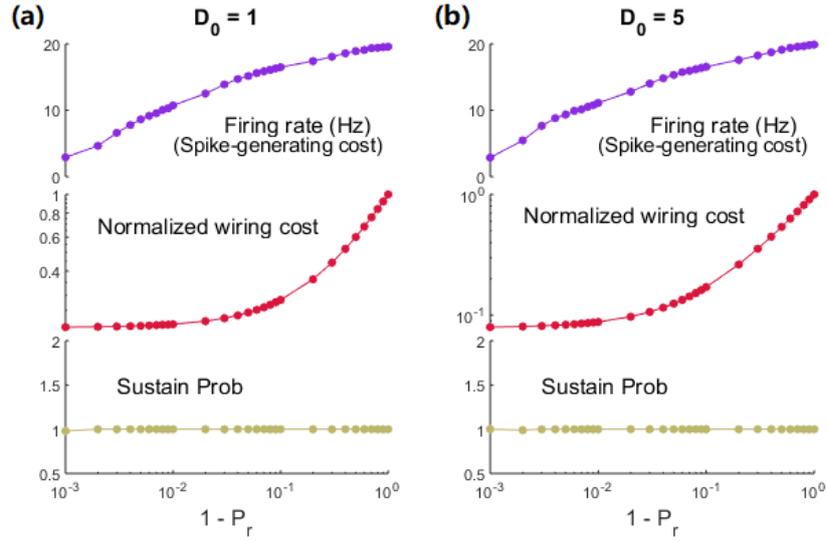

Fig. S6. Results of an extended model setting where the number of inter-modular links depends on the distance between modules. Here, the distance between two modules $k$ and $l$, denoted as $D_{kl}$, is defined as the distance between the center position of two modules and the number of inter-module links between two modules is proportional to $e^{-D_{kl}/D_0}$, where $D_0$ is a parameter which determines the characteristic length. From top to bottom: The firing rate, normalized wiring, and the sustained probability versus the rewiring probability are presented. $D_0 = 1$ and $D_0 = 5$ in (a) and (b) respectively. The network size is $N = 50{,}000$ with density $P_c = 0.0017$.

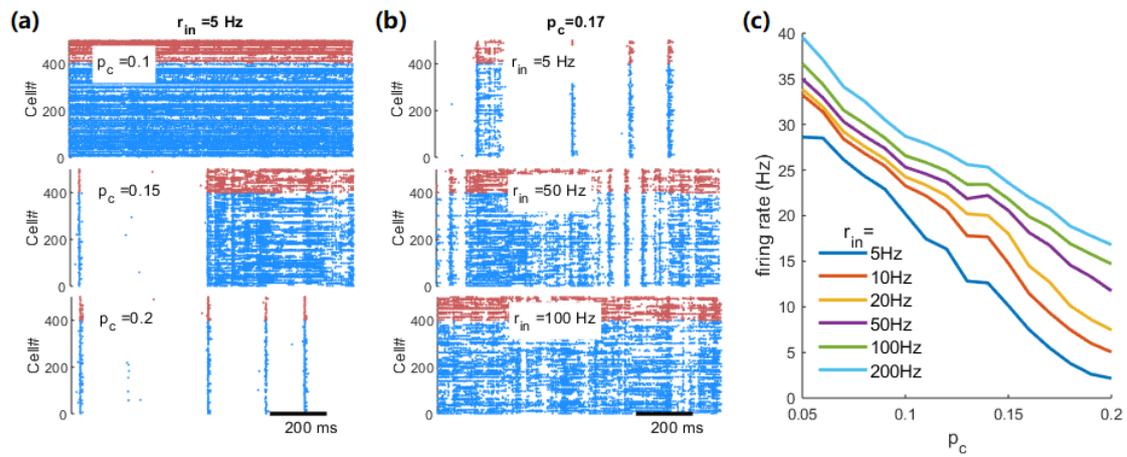

Fig. S7. Additional simulation of single separate module with different density $p_c$ and external input rate $r_{in}$. (a) The raster plot of spikes with $r_{in} = 5$ Hz and connection densities $p_c = 0.10$, 0.17, and 0.20 respectively. Under fixed input the dynamic transitions from asynchronous spiking to synchronous spiking with decreased rate. (b) The raster plot of spikes with density $p_c = 0.17$ and $r_{in} = 5$, 50, and 100 Hz respectively. Under fixed connection density, the dynamic transitions from synchronous spiking to asynchronous spiking as with increased input rates. (c) The firing rate with different density $p_c$ and input rate $r_{in}$. The firing rate of the network always decreases with the connection density under various input rates.

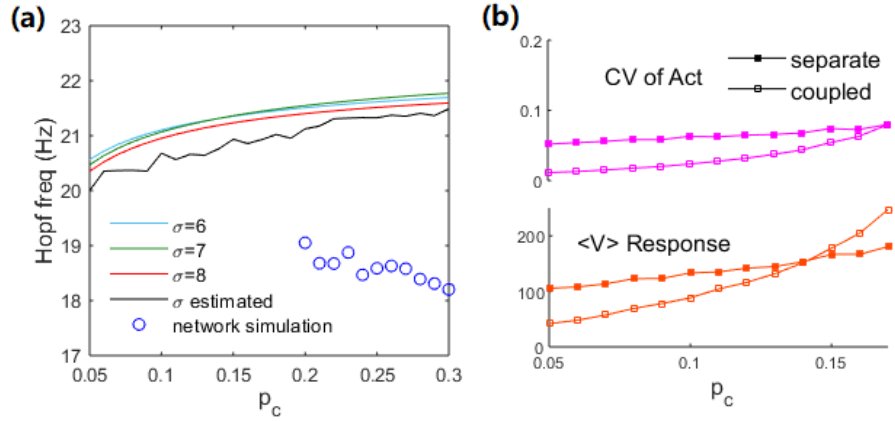

Fig. S8. Additional comparison of mean-field theory prediction. (a) The Hopf frequency, defined by imaginary part divided by $2\pi$ of the dominant eigenvalue of the fixed point, of the single-module field equation Eq. (2). Different curves are results by fixing $\sigma_\alpha = 6,7,8$, and by 'optimal' $\sigma_\alpha$ given in Fig. 4(a). Blue circles are peak frequencies of $V_E$ oscillation from network simulation. Since the single module system approaches but has not reached the Hopf bifurcation point when increasing $p_c$ (Fig. 6(b)), together with the finite size effect from the small number of neurons in a module, the prediction of oscillatory frequency is not very precise with errors around a few Hz. (b) The comparison of the CV of activity and response size of membrane potential between 1) the single-module field equations (Eq. (2)) with different density $p_c$ (solid markers, results in Fig. 6(b)) and 2) the coupled-modules field equations (Eq. (5)) with corresponding modular density $p_c$ (hollow markers, transferred by results in Fig. 6(c) through $p_c = P_c(1 + (N_m - 1)P_r)$).